\documentclass[journal]{IEEEtran}

\IEEEoverridecommandlockouts
\usepackage{cite}
\usepackage{amsmath,amssymb,amsfonts}
\usepackage{algorithmic}
\usepackage{graphicx}
\usepackage{textcomp}
\usepackage{xcolor}
\usepackage{tabularx}
\usepackage{pgfplots}
\pgfplotsset{compat=1.5}
\usepackage{pgfplotstable}
\usepgfplotslibrary{statistics}
\pgfplotsset{grid style={dotted,gray}}
\usepackage{tikz}
\definecolor{bblue}{HTML}{4F81BD}
\usepackage{soul}

\DeclareMathOperator{\E}{\mathbb{E}}
\def\BibTeX{{\rm B\kern-.05em{\sc i\kern-.025em b}\kern-.08em
    T\kern-.1667em\lower.7ex\hbox{E}\kern-.125emX}}
\begin{document}

\title{Slotted Aloha with Capture for OWC-based IoT: Finite Block-Length Performance Analysis\\
\author{Tijana Devaja,~\IEEEmembership{Student Member,~IEEE,} Milica Petkovic,~\IEEEmembership{Member,~IEEE,} Francisco J. Escribano,\\
        ~\IEEEmembership{Senior Member,~IEEE,} \v Cedomir Stefanovi\' c,~\IEEEmembership{Senior Member,~IEEE,} Dejan Vukobratovic,~\IEEEmembership{Senior Member,~IEEE}}
\thanks{This paper was presented in part at the 17th International Symposium On Wireless Communication Systems (ISWCS) 2021 \cite{9562172}.}
\thanks{T.~Devaja, M.~Petkovic and D. Vukobratovic are with University of Novi Sad, Faculty of Technical Sciences, 21000 Novi Sad, Serbia (e-mails: tijana.devaja@uns.ac.rs; milica.petkovic@uns.ac.rs;  dejanv@uns.ac.rs).}
\thanks{F.~J.~Escribano is with Universidad de Alcal\'{a}, Alcal\'{a} de Henares, Spain (e-mail: francisco.escribano@uah.es).}
\thanks{\v C.~Stefanovi\' c is with Aalborg University, Aalborg, Denmark (e-mail: cs@es.aau.dk).}
}

\makeatletter
\newcommand{\linebreakand}{%
  \end{@IEEEauthorhalign}
  \hfill\mbox{}\par
  \mbox{}\hfill\begin{@IEEEauthorhalign}
}
\makeatother

\author{
}

\maketitle

\begin{abstract}

In this paper, we propose a Slotted ALOHA (SA)-inspired solution for an indoor optical wireless communication (OWC)-based Internet of Things (IoT) system containing IoT devices that exchange data with an access point (AP).
Assuming that the OWC receiver at the AP exploits the capture effect, we derive % are interested in the derivation of 
the error probability of decoding a short-length data packet originating from a randomly selected OWC IoT device in the presence of %a given number of 
interfering users.
The analysis rest on the derivation of the signal-to-noise-and-interference-ratio (SINR) statistics and the application of the finite block-length (FBL) information theory.
Using the analytical results, we derive meaningful performance parameters such as system throughput and reliability (expressed in terms of the outage probability of a user transmission). The main trade-offs between the system performance and the OWC system setup parameters are investigated, stressing how the indoor OWC-based system geometry plays an important role in the system performance. Using extensive numerical results, we demonstrate how they can serve as a guideline for a system design aiming at the optimization of the proposed SA-based random access protocol.

\end{abstract}


\section{Introduction}

Optical wireless communications (OWC) have been recently gaining attention as a technology able to offload traffic from wireless radio-frequency (RF) technologies in indoor environments.
This may alleviate the challenges posed by the spectrum crunch that affects the most intensively exploited bands. %RF spectrum.
In fact, the RF spectrum shortage is continuously pushing for the opening of higher frequency bands for wireless communications, ranging from millimeter-wave, over terahertz to optical (infrared (IR) or visible) bands~\cite{ref1, OWC_MATLAB, ref9, JSAC}.
Each of these possibilities target specific environments and applications, and the ongoing research points towards a combination of access strategies to meet the ever-increasing  Internet of Things (IoT) requirements and the associated demand for resources.

In any case, using a specific frequency band for an IoT application is only a part of a wider context.
%If we focus on the needs of
Specifically, in a wireless IoT framework, where the communication medium is shared, granting a fair access to the shared resources and guaranteeing the link quality in presence of possible interference is of paramount importance~\cite{PP,modernRA}, while the sporadic nature of the IoT traffic and the short lengths of the exchanged data packets pose specific challenges in this regard. % that are co randomly and without centralized coordination, only adds to the challenges to be addressed.
In this context, well designed random access (RA) protocols are instrumental to obtain a good performance of the access network~\cite{RA0,RA1,RA2,RA3,Jeon22, Grybosi22}.
Along the last decades, there has been a general trend to embrace interference in wireless media in order to efficiently exploit the available, scarce spectrum resources (e.g. by using spread spectrum or MIMO spatial multiplexing techniques).
In the domain of access schemes, the same principle is applied through exploitation of the capture effect, which frequently occurs in wireless communications and through which user data may be recovered %under specific conditions even 
in presence of concurrent transmissions.
Under such scenarios, collision avoidance ceases to be one the basic design principles for successful RA protocols, because, in fact, promoting collisions may become beneficial.
%This removes the need to  collision-avoidance RA protocols,  promoting simple and efficient RA schemes inspired by the ALOHA protocol.
In this way, a simple slotted ALOHA (SA) solution in combination with the exploitation of the capture effect at the physical layer can achieve a favorable throughput performance~\cite{ref4,ref5,ref6}.

The IoT deployment scenario considered in this paper assumes an indoor space comprising large and unobstructed areas (e.g., a warehouse or an open-plan office), where a large quantity of devices collect, process and send data to a number of access points.
In such situation, the use of RF links could lead to an intolerable level of interference and a high probability of packet loss due to collisions.
However, transmission technologies based on OWC can help to tackle these challenges, despite their limited coverage.
In fact, by resorting to OWC links, the interference across walls is eliminated, and a small scale indoor cellular network could be designed to cover the whole space with a minimal coverage overlapping, thus maximizing the overall throughput.

OWC systems were studied in the recent literature as a potential indoor IoT solution in current and upcoming generations of mobile communication technologies~\cite{b10,0,2, OWCindoor1,OWCindoor2,OWCindoor2a,OWCindoor3,Wu20}.
For the interested reader, a detailed state-of-the-art of the  OWC for the IoT has been presented in \cite{SurveyOWC}, mostly focusing on four main IoT domains: Internet of Terrestrial Things (IoTT), Internet of underWater Things (IoWT), Internet of Biomedical Things (IoBT), and Internet of underGround Things (IoGT).
Different uplink RA approaches with sporadic and varying device activity in OWC-based IoT systems were analyzed in \cite{OWCindoor5, OWCindoor5a, OWCindoor5b, OWCindoor5c}.
Specifically, \cite{OWCindoor5} analyzed the uplink multi-receiver OWC system in the context of a massive IoT application, where a coded SA approach with successive interference cancellation was adopted. 
In \cite{OWCindoor5b}, Zhao \textit{et al.} analyzed a Multi-Packet Reception (MPR)-aided Visible Light Communications (VLC) system and introduced a novel Quality of Service (QoS)-driven non-carrier sensing RA scheme.
The same authors also proposed a novel SA algorithm with heterogeneous delay that guarantees the QoS in the context of the MPR VLC system~\cite{OWCindoor5a}.
An optical camera communication based on the ALOHA RA scheme was assessed in \cite{OWCindoor5c}, where the access probability of each terminal was optimized in order to guarantee access fairness and system throughput rate maximization.

In this paper, we push the state-of-the-art by analyzing and optimizing a slotted ALOHA-based access scheme that takes into account an advanced modelling of the physical layer via (i) characterization of the signal-to-interference ratio (SINR) in the presence of interfering, randomly activated users and (ii) characterization of the impact of the finite-block length (FBL) effects.
The obtained analytical framework allows the derivation of the probability that a user transmission is decoded in the presence of the rest of interfering users. Moreover, other performance metrics can be derived and the access scheme optimized.
To the best of our knowledge, this is the first work that investigates the FBL effects in such an OWC environment.
%The design and optimization of an RA scheme based on the capture effect for an OWC-based IoT indoor network calls for the characterization of the signal-to-interference ratio (SINR) in the presence of interfering, randomly activated users.
%Another key component to be taken into the analysis is the impact of the channel code which protects the transmitted packets.
%From the SINR statistics, one can derive the error probability affecting the transmitted data packets, protected by an appropriate channel code %thereby determining the system performance.}
%However, consideration of the impact of channel code in an OWC environment has not yet been treated in the literature to the best of our knowledge.
%and how it is affected by the meaningful parameters of the system, a deep error and coding rate analysis is still lacking in open literature.
In particular, information theory has traditionally focused on data transmission reliability through noisy channels that may be achievable for channel coded data with large block lengths, but the fact is that, for IoT applications, we are usually dealing with short block-length transmissions.
In this sense, the so-called FBL information theory~\cite{Polyanskiy} provides us with the appropriate tools to analyze what relates to error probability estimation and the corresponding maximum achievable rate under a given message length constraint.
%The aim of this paper is a detailed analysis of the OWC-based IoT system performance in FBL regime. 

The presented work is an extension of the initial study of the SINR statistics for an OWC-based IoT system with capture effect presented in~\cite{9562172}.
In this respect, the main novelty of the paper is the error probability analysis based on FBL theory, as well as the analysis of the overall system throughput and the reliability in terms of the outage probability.
%\CS{The aim  of this study is to derive SINR for an OWC-based IoT system with capture effect. - I miss here some kind of account what is novel wrt the ISWCS work.}
%Its main novelty is the development of the error probability analysis based on FBL theory, as well as the analysis of the overall system throughput and the reliability in terms of the outage probability.
%Moreover, the evaluation of the impact on the overall system performance of the main parameters of the OWC-based IoT framework is presented.
The FBL performance analysis for this kind of OWC-based IoT system represents the appropriate method to set the system parameters during the system design, with the aim to achieve an optimal performance of the RA protocol. 

The contributions of the paper can be summarized as follows:

1) We propose an architecture for future IoT based on short-range OWC technology and design an access scheme %with the capture effect 
for an OWC-based IoT indoor network. %, is performed based on the the analysis and evaluation 
We analyze the scheme in the terms of SINR of the accessing users and evaluate its performance. 
The analysis builds up on our previous work \cite{9562172}, %we have partially characterized the SINR in this environment and how it is affected by the parameters of the system, but 
which is here complemented by the error and coding rate characterization, enabling an accurate modelling of the physical layer effects. 

2) The error and the coding rate characterizations are based on the FBL performance analysis for OWC-based IoT systems that is exposed in the paper. Specifically, from SINR statistics it is possible to derive the error probability affecting the captured data packets, protected by an appropriate channel code, thereby determining the system performance.

3) %Besides the error probability analysis based on FBL theory, the main novelty of 
 Building up on the physical layer characterization, the paper presents the analysis of the overall system throughput and the reliability in terms of the outage probability. 

4) We use numerical results to provide guidelines for the optimal system design, %The optimal design of indoor OWC systems is discussed 
by exploring the effects of OWC parameters on the system performance. These contributions are particularly important, as a real-world IoT network deployment requires not only that its physical parameters are appropriately set, but also that the access protocol is properly configured to achieve the optimal performance.

The rest of the paper is structured as follows.
In Section II, we review the system model and set the main hypothesis.
In Section III, we derive the SINR statistics for the proposed setup.
In Section IV, we analyze the error probability and channel coding rate based on the FBL regime approach, as well as the overall system throughput and the reliability in terms of the outage probability.
In Section V, we present and discuss the numerical results and, finally, in Section VI, we close the paper with the main conclusions.

\section{SA-based Indoor OWC IoT: System Model}

The context of this work comprises a communication scenario in which a total of $U$ IoT devices equipped with OWC transmitters contend to access a common OWC access point (AP).
The transmitting devices are uniformly placed on a horizontal plane, while the OWC AP is located at the ceiling, at a fixed location (see Fig.~\ref{Fig1}).
This setup corresponds to an open-plan office space with IoT devices placed on tables or other furniture.
The SA protocol~\cite{RA2} is used for uplink transmissions. %, using slots that accommodate a single packet transmission. %of duration \CS{$T_\text{sl}$}.
Each IoT device is active for transmission with probability $p_a$ in every slot, independently of its activity in other slots and of the activity of other devices during the same slot period.
In other words, we consider Bernoulli arrivals on user basis.
If a user is active in a slot, it transmits a fixed-length packet with finite (short) block-length.
We denote the set of active users in a slot by $\mathcal{U}_a$, and by $U_a = |\mathcal{U}_a|$ the corresponding number of active users, where $0\leq U_a \leq U$.
It is straightforward to show that $U_a$ is a random, binomially distributed parameter $\mathcal{B}(U, p_a)$.
Finally, with a slight misuse of notation, we denote the individual active users by $u_i \in \mathcal{U}_a$, $i = 1, \cdots, U_a$, where their indexing is assigned arbitrarily.
%We will derive expressions in our analysis as a function of $U_a$, and then provide averages over it.
%In the sequel, we assume Bernoulli arrivals for the activation of the IoT devices, in order to model the distribution of $U_a$.
%\CS{, can be modeled by the binomial distribution, i.e., $U_a=\mathcal{B}(U, p_a)$}.
%The average number of packets sent per slot is defined as the channel load $G$ [pk/slot], and is equal to
%\begin{equation}
%G =  \mathbb{E} \left[ U_a \right] = p_a U.
%\label{G}
%\end{equation}

\begin{figure}[b!]
\centerline{\includegraphics[width=0.97\columnwidth]{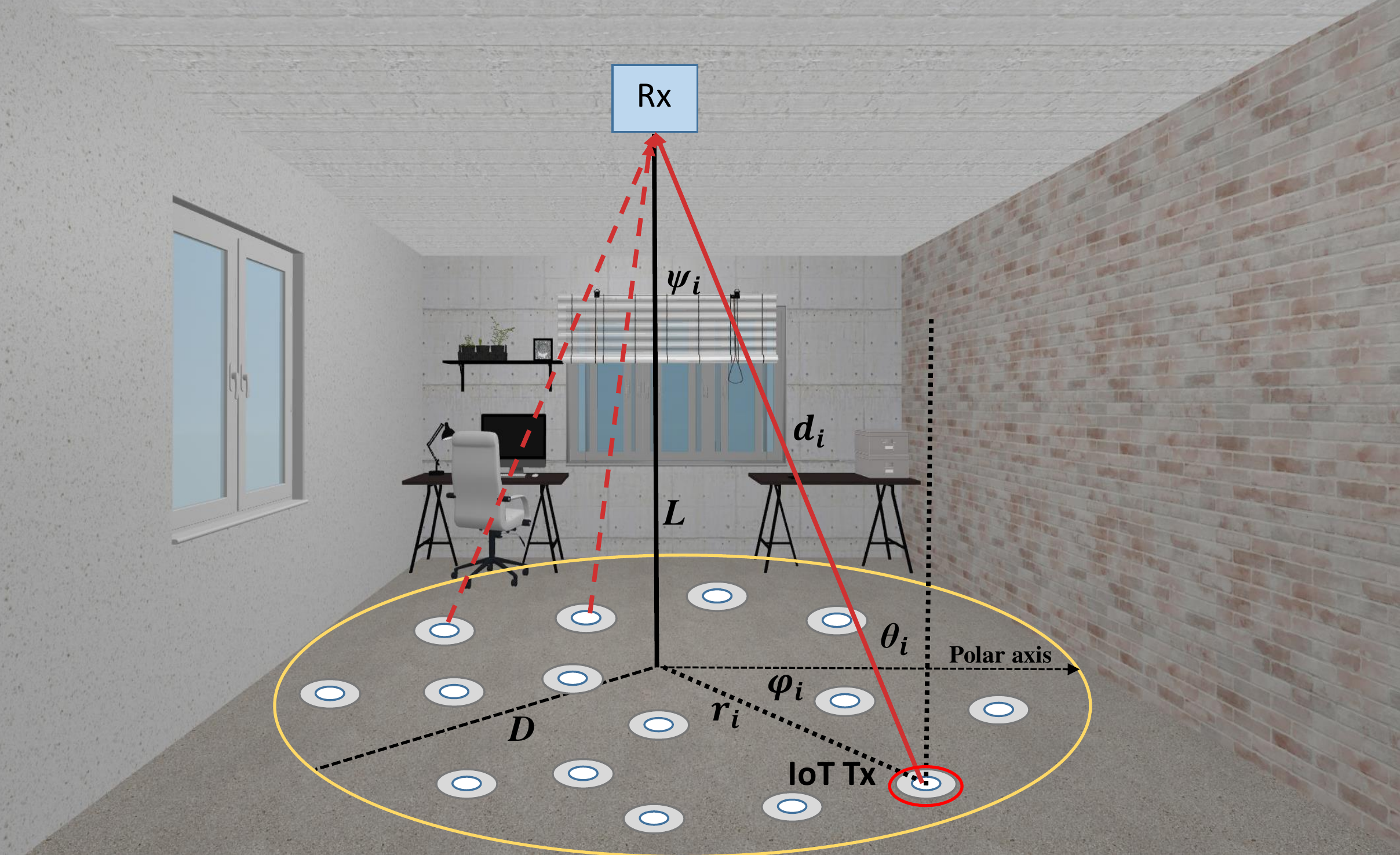}}
\caption{OWC-based indoor IoT system model.}
\label{Fig1}
\end{figure}

The IoT devices use IR LED sources, and the physical transmission employs a simple intensity-modulation (IM) binary-format signaling (e.g., non-return-to-zero on-off keying, OOK).
All  devices transmit with the same optical power regardless of any other condition.
The OWC photo-detector (PD) receiver performs direct detection (DD) of the arriving light intensity~\cite{OWC_MATLAB}.
At any given slot, the light intensity impinging the OWC PD receiver comprises the contribution of the $U_a$ active users plus the background radiation noise.
Assuming that the PD is working in a linear regime (which is a standard and a reasonable approximation in practical situations), the received signal after the conversion to the electrical domain (prior to demodulation and decoding) can be modeled as
\begin{equation}
\begin{split}
y(t)= \sum_{i=1}^{U_a} P_t \eta h_i x_i(t) + n(t),
\end{split}
\label{y1}
\end{equation}
where $x_i(t)$, $i=1, \cdots, U_a$, is the unit-power signal waveform of the active user $u_i$, $P_t$ is the transmitted optical power, $\eta$ the optical-to-electrical conversion coefficient, $h_i \geq 0$ the optical channel gain from the user $u_i$ to the AP, while $n(t)$ is an instance of additive white Gaussian noise with power spectral density $N_0/2$.
This kind of noise process adequately models the distorting effects of the optical background radiation and the receiver thermal noise, and contributes to the overall  signal-to-noise  with a noise power $\sigma^2_n=N_0 B$, where $B$ is the system noise bandwidth.

Under these conditions, if a slot contains transmissions from two or more users (i.e., $U_a \geq 2$ for such slot), a packet collision takes place.
In that case, by exploiting the capture effect, the OWC AP receiver will try to decode the colliding transmissions, in contrast to the classical SA scheme, which assumes the collision channel model.
We are interested in the probability that the receiver succeeds in decoding a randomly selected user among the set of active ones.
For convenience, and without loss of generality, we assume that the index of such user (denoted as the {\it reference user} henceforth) corresponds to $i = 1$, i.e., $u_1$ is the reference user, while the rest of the active users $u_i$, $i=2, \cdots, U_a$, are the interfering ones.
Therefore, the received signal (\ref{y1}) can be rewritten as
\begin{equation}
\begin{split}
y(t)= P_t \eta h_{1} x_{1}(t) + \sum_{i=2}^{U_a} P_t \eta h_i x_i(t) + n(t),
\end{split}
\label{y2}
\end{equation}
where $x_{1}(t)$ is the signal waveform from the reference user $u_1$, $h_{1} \geq 0$ is the corresponding optical channel gain, and the summation term represents the interference contribution from all other active users.

Based on \eqref{y2}, the SINR experienced by the reference user can be readily written as
\begin{equation}
% \begin{split}
% {\rm SINR} & = \frac{ P_t^2 \eta^2 h_{1}^2 }{ \sum_{i=2}^{U_a}P_t^2 \eta^2 h_i^2 + \sigma _n^2} \\
% & = \frac{\gamma_{1}  }{ \sum_{i=2}^{U_a} \gamma_i + 1} = \frac{\gamma_{1}  }{ \gamma_{\rm I} + 1},
% \label{sinr}
% \end{split}
{\rm SINR} = \frac{ P_t^2 \eta^2 h_{1}^2 }{ \sum_{i=2}^{U_a}P_t^2 \eta^2 h_i^2 + \sigma _n^2} = \frac{\gamma_{1}  }{ \sum_{i=2}^{U_a} \gamma_i + 1} = \frac{\gamma_{1}  }{ \gamma_{\rm I} + 1}
\label{sinr},
\end{equation}
where 
\begin{equation}
\gamma_{1}  = \frac{ P_t^2 \eta^2   h_{1}^2 }{\sigma _n^2},~\gamma_i  = \frac{ P_t^2 \eta^2 h_i^2}{\sigma _n^2},~\gamma_{\rm I}  =\!\! \sum_{i=2}^{U_a} \!\!\gamma_i.
\label{gUa}
\end{equation}
The SINR depends not only on the number of active users, but also on the specific placements of the IoT devices with respect to the OWC AP (which, in turn, determines the specific values of the optical path gains, $h_i$). In the next section, we derive the statistics for the SINR observed by a randomly selected active user, based on the characteristics of the system setup and its optical parameters.

\section{SINR Statistics for the SA-based OWC IoT system}
 
%\begin{figure}[!b]
%\centering
%\includegraphics[width=2.8in]{Fig2}
%\caption{Geometry of the  propagation model.}
%\label{Fig2}
%\end{figure}

As shown in Fig.~\ref{Fig1}, we assume that the devices are randomly and uniformly distributed over the same horizontal plane within the indoor circular coverage area of radius $D$.
The OWC AP is positioned at a height $L$ above the plane, over the center of the circle.
The location of user $u_i$ with respect to the OWC AP is determined by the angle $\theta_i$ (angle of irradiance), the angle $\varphi_i$, and the radius $r_i$ in the polar coordinate plane.
Furthermore, the corresponding angle of incidence into the AP is denoted as $\psi_i$, and the Euclidean distance between the corresponding LED transmitter and the PD receiver is denoted as $d_i$.
The optical channel gain of the line-of-sight (LoS) link between $u_i$ and the PD receiver can be determined as~\cite{IR}
\begin{equation}
h_i = \frac{ A_r\left( m + 1\right) R_r}{2\pi d_i^2}\cos ^m\left( \theta_i \right)T_s g \left( \psi_i \right)  \cos \left( \psi_i \right),
\label{I_n1}
\end{equation}
where $ A_r $ is the surface area of the PD, $R_r$ is the responsivity, $T_s$ is the gain of the optical filter, $g\left( \psi_i \right)$ is the response of the optical concentrator, and the factor $m$ is described in the sequel.
The optical concentrator is modeled as $g\left( \psi_i \right) =\zeta^2 /\sin^2\left( \Psi \right)$, for $0\leq\psi_i \leq \Psi$, where $\zeta$ is the refractive index of the lens at the PD and $\Psi$ denotes its field of view (FoV).
Finally, the LED emission follows a generalized Lambertian radiation pattern with order $m = -\ln 2/ \ln \left( \cos \Phi_{1/2} \right)$, where $\Phi_{1/2}$ denotes the semi-angle at half illuminance\footnote{We assume that all LEDs are characterized by the same parameters, i.e., $m_i=m$ and $\Phi^i_{1/2}=\Phi_{1/2}$ $\forall i$.}~\cite{OWC_MATLAB}.

In this situation, if the surface of the PD receiver is parallel to the plane where the IoT devices are located, then $\theta_i =\psi_i $, $d_i=~\sqrt {r_i^2 + L^2} $,    $\cos \left( \theta_i \right) \!=\! \frac{L}{ \sqrt {r_i^2 + L^2} }$, and equation (\ref{I_n1}) can be rewritten as
\begin{equation}
h_i = \frac{\mathcal X  }{\left( r_i^2 + L^2 \right)^{\frac{m + 3}{2}}},
\label{I_n2}
\end{equation}
where $\mathcal X = \frac{A_r\left( m + 1 \right)R_r}{2\pi}T_s g\left( \psi_i \right)L^{m + 1}$ is a factor that does not depend on the specific placement of the IoT device, provided that the angle $\psi_i$ lies below the PD FoV (a condition met for the values of $D$ and $L$ considered here, so that there would be no IoT devices excluded from the coverage area of the AP).

As the IoT devices are uniformly distributed, the PDF of their radial distance $r_i$ from the centre of the circle is \cite{SA3}
\begin{equation}
f_{r_i}\left( r \right) = \frac{2r}{D^2},\quad 0\leq r \leq D.
\label{pdf_rn}
\end{equation}
By using expressions \eqref{I_n2} and \eqref{pdf_rn}, and standard RV transformation techniques, the PDF of the optical channel gains, $h_i$, can be derived as
\begin{equation}
f_{h_i}\left( h \right) = \frac{2 \mathcal X ^{\frac{2}{m + 3}}}{D^2\left( m + 3 \right)}  h^{ - \frac{m + 5}{m + 3}},\quad  h_{\min }\leq h \leq  h_{\max },
\label{pdf_In}
\end{equation}
where $ h_{\min} = \frac{\mathcal X }{{\left( D^2 + L^2 \right)}^{\frac{m + 3}{2}}}$ and  $h_{\max} = \frac{\mathcal X}{L^{m + 3}}$.

The PDF of $\gamma_i$, defined in \eqref{gUa}, can be similarly derived as~\cite{SA3}
\begin{equation}
f_{\gamma_i}\left( \gamma  \right) = \frac{(\mu\mathcal X^2)^{\frac{1}{m + 3}}}{D^2\left( m + 3 \right)}\gamma^{ - \frac{m + 4}{m + 3}},\quad \gamma_{\min }\leq \gamma \leq \gamma_{\max},
\label{pdf_hi2}
\end{equation}
where $\gamma_{\min} = \frac{\mu \mathcal X^2}{{\left(D^2 + L^2 \right)}^{m+ 3}}$,   $\gamma_{\max} = \frac{ \mu \mathcal X^2}{L^{2 \left( m+ 3\right)}}$, and $\mu= \frac{ P_t^2 \eta^2 }{\sigma _n^2}$.

The CDF of $\gamma_i $ can be easily derived as
\begin{equation}
F_{\gamma_{i}}\left( \gamma  \right) = 
\begin{cases}
%1+ \frac{L^2}{R^2} - \frac{(\mu\mathcal X^2) ^{  \frac{1}{m + 3}}}{R^2}\gamma^{- \frac{1}{m + 3}}, &  \gamma_{\min}\leq \gamma \leq\gamma_{\max}  \\
1+ \frac{L^2 - (\mu\mathcal X^2 / \gamma) ^{  \frac{1}{m + 3}} }{R^2}, &  \gamma_{\min}\leq \gamma \leq\gamma_{\max} \\
1, &  \gamma > \gamma_{\max}
\end{cases}.
\label{cdfVLC}
\end{equation}

To get the statistical characterization of the overall SINR, we can use the characteristic function (CF) approach~\cite{Papoulis}. In this case, the CF of $\gamma_i = \frac{ P_t^2 \eta^2  h_i^2}{\sigma _n^2}$ can be derived via (\ref{pdf_hi2}) as
\begin{equation}
\begin{split}
& \varphi_{\gamma_i} \left( t \right)  \triangleq  \E \left[ e^{jt\gamma_i} \right] = \int_{-\infty}^{\infty} e^{jtx}  f_{\gamma_i}\left( \gamma  \right) \,\mathrm{d}\gamma \\
& = \frac{(\mu\mathcal X^2)^{\frac{1}{m + 3}}}{D^2\left( m + 3 \right)} \int_{\gamma_{\min}}^{\gamma_{\max}} \gamma^{ - \frac{m + 4}{m + 3}} e^{jt\gamma}   \,\mathrm{d}\gamma  = \frac{(\mu\mathcal X^2)^{\frac{1}{m + 3}}}{D^2\left( m + 3 \right)} \\
& \times\left( \Gamma\!\left(-\frac{1}{m+3},-j t \gamma_{\rm min}  \right) \!\! -\! \Gamma\! \left(-\frac{1}{m+3}, -j t \gamma_{\rm max}\!  \right) \!\! \right)\!\!,
\end{split}
\label{cf_gi}
\end{equation}
where $\Gamma\left(s, z \right) = \int_z^\infty t^{s-1}e^{-t} \mathrm{d}t  $ is the upper incomplete gamma function \cite[(8.35)]{grad}.
%\begin{equation}
%\begin{split}
%& \varphi_{\gamma_i} \left( t \right)  \triangleq  \E \left[ e^{it\gamma_i} \right] = \int_{-\infty}^{\infty} e^{it\gamma}  f_{\gamma_i}\left( \gamma  \right) \,\mathrm{d}\gamma = \frac{(\mu\mathcal X^2)^{\frac{1}{m + 3}}}{R^2\left( m + 3 \right)} \int_{\gamma_{\min}}^{\gamma_{\max}} \gamma^{ - \frac{m + 4}{m + 3}} e^{it\gamma}   \,\mathrm{d}\gamma \\
%& = \frac{(\mu\mathcal X^2)^{\frac{1}{m + 3}}}{R^2\left( m + 3 \right)} 
%\left( \Gamma \left(-\frac{1}{m+3},-i t \gamma_{\rm min}  \right)  - \Gamma \left(-\frac{1}{m+3}, -i t \gamma_{\rm max}  \right)  \right) \\
%& = \frac{(\mu\mathcal X^2)^{\frac{1}{m + 3}}}{R^2\left( m + 3 \right)}
%\left( \gamma_{\rm min}^{-\frac{1}{m+3}} E_{\frac{m+4}{m + 3}} \left( -it \gamma_{\rm min} \right) - \gamma_{\rm max}^{-\frac{1}{m+3}} E_{\frac{m+4}{m + 3}} \left(-it \gamma_{\rm max} \right)  \right).
%\end{split}
%\label{cf_gi}
%\end{equation}

\subsection{Contribution to the SINR Statistics from the Reference User}% ($\gamma_{U_a}$)}

For the reference user, the PDF of $\gamma_{1}$ can be calculated as a particular case of equation (\ref{pdf_hi2}), rewritten here for clarity
\begin{equation}
f_{\gamma_{1}}\left( \gamma  \right) = \frac{(\mu\mathcal X^2)^{\frac{1}{m + 3}}}{D^2\left( m + 3 \right)}\gamma^{ - \frac{m + 4}{m + 3}},\quad \gamma_{\min }\leq \gamma \leq \gamma_{\max }.
\label{pdf_gUa}
\end{equation}

\subsection{Contribution to the SINR Statistics from the Interfering Users} %($\gamma_{\rm I}$)}

The channel gains $h_i$ (thus also $\gamma_i$) are independent and identically distributed (i.i.d.) RVs when the IoT device locations are also i.i.d. RVs. In this case, the CF of $\gamma_{\rm I} = \sum_{i=2}^{U_a} \gamma_i$ can be determined as~\cite{Papoulis}
\begin{equation}
\begin{split}
 \varphi_{\gamma_{\rm I}} \left( t \right) & \triangleq  \E \left[ e^{jt\gamma_{\rm I}} \right] = \E \left[ e^{jt \sum_{i=2}^{U_a} \gamma_i } \right] = \E \left[ \prod_{i=2}^{U_a} e^{jt \gamma_i  } \right] \\
& =  \prod_{i=2}^{U_a}  \E \left[ e^{jt \gamma_i  } \right]  =  \prod_{i=2}^{U_a} \varphi_{\gamma_i} \left( t \right) =   \varphi_{\gamma_i}^{U_a-1}  \left( t \right),
\end{split}
\label{cf_isi}
\end{equation}
where the CF of $\gamma_i$ is defined in (\ref{cf_gi}). The PDF of $\gamma_{\rm I}$ can be derived as
\begin{equation}
\begin{split}
 f_{\gamma_{\rm I}} \left( \gamma \right) &  = \frac{1}{2\pi} \int_{-\infty}^{\infty} e^{-jt\gamma}  \varphi_{\gamma_{\rm I}} \left( t \right) \,\mathrm{d}t  \\
& = \frac{1}{2\pi} \int_{-\infty}^{\infty} e^{-jt\gamma}  \varphi_{\gamma_i}^{U_a-1}  \left( t \right) \,\mathrm{d}t,
\end{split}
\label{pdf_isi}
\end{equation}
for $\gamma_{\min }^{U_a-1}\leq \gamma \leq \gamma_{\max }^{U_a-1}$.

\subsection{Overall SINR Statistics for the Reference User}% ($\frac{\gamma_{U_a}  }{ \gamma_{\rm I} + 1}$)}

As an intermediate step, we derive the PDF of the RV defined as $\lambda = \gamma_{\rm I} + 1$, namely
\begin{equation}
\begin{split}
 f_{\lambda} \left( \lambda \right)  &  =  \frac{f_{\gamma_{\rm I}}\left(  \gamma_{\rm I} \right)}{\vert \frac{\mathrm{d}\lambda}{\mathrm{d} \gamma_{\rm I}} \vert } = f_{\gamma_{\rm I}} \left( \lambda -1 \right)  \\
 & = \frac{1}{2\pi} \int_{-\infty}^{\infty} e^{-jt\left( \lambda -1 \right)}  \varphi_{\gamma_i}^{U_a-1}  \left( t \right) \,\mathrm{d}t ,
\end{split}
\label{pdf_isi1}
\end{equation}
for $\gamma_{\min }^{U_a-1} + 1\leq \lambda \leq \gamma_{\max }^{U_a-1}+ 1$.  
Since {$ \gamma_{1} $ and $ \gamma_{\rm I} $ are independent RVs, their joint PDF is $f_{\gamma_{1},\lambda} \left( x,\lambda \right)=f_{\gamma_{1}} \left( x \right)f_{\lambda} \left( \lambda \right)$}.
The PDF of the ${\rm SINR} =\frac{\gamma_{1}}{\lambda}=\frac{\gamma_{1}}{\gamma_{\rm I} + 1}$ of the reference user, conditioned on the total number of active users $U_a$, can be written %  by applying  the technique for transformation of random variables,
as follows~\cite{ratio}
\begin{equation}
\begin{split}
 f_{\rm SINR} \left( x | U_a \right) = \int_{-\infty}^{\infty}\vert\lambda \vert  f_{\gamma_{1}} \left(x\lambda \right) f_{\lambda} \left( \lambda \right) \mathrm{d}\lambda,
\end{split}
\label{pdf_sinr}
\end{equation}
where $f_{\gamma_{1}} \left(x \right)$ and $f_{\lambda } \left(\lambda  \right)$ are defined in equations (\ref{pdf_gUa}) and (\ref{pdf_isi1}), respectively.
After replacing (\ref{pdf_gUa}) and (\ref{pdf_isi1}) in equation (\ref{pdf_sinr}), since $\lambda>0$, this PDF can be written as
\begin{equation}
\begin{split}
 &  f_{\rm SINR}\left( x | U_a \right)  = \frac{(\mu\mathcal X^2)^{\frac{1}{m + 3}}}{2\pi D^2\left( m + 3 \right)} x^{ - \frac{m + 4}{m + 3}} \\
 & \!\times \int_{\gamma_{\min }^{U_a-1} + 1}^{\gamma_{\max }^{U_a-1} + 1}\!\!  \lambda ^{ - \frac{1}{m + 3}}  \left( \int_{-\infty}^{\infty} \!\! \!e^{-jt\left( \lambda -1 \right)}  \varphi_{\gamma_i}^{U_a-1}  \left( t \right) \,\mathrm{d}t \right) \mathrm{d}\lambda,
\end{split}
\label{pdf_sinr1}
\end{equation}
where $\varphi_{\gamma_i}\left( t \right)$ is the CF previously defined in equation (\ref{cf_gi}).

Finally, the CDF of the ${\rm SINR}$ of the reference user, conditioned on the number of active users in the slot $U_a$, can be written as
\begin{equation}
\begin{split}
F_{{\rm SINR}}\left( \gamma  \right | U_a )= \int_{0}^{\gamma}  f_{\rm SINR } \left( x | U_a \right) \,\mathrm{d}x 
\end{split}.
\label{cdf_sinr}
\end{equation}

This statistical characterization of the SINR will allow us to derive the error probability in the next section.

\section{Performance Analysis for the OWC-based IoT System based on SA via the FBL approach}

\subsection{Error Probability for SA via the FBL approach}

The basic goal of information theory is to quantify the extent to which reliable communication is possible over a noisy channel. A code of size \textit{N} and block length \textit{n} allows the communication of one of \textit{N} messages via \textit{n} uses of the channel. The fundamental trade-off between these quantities and the reliability of the communication process is captured by $N_{\epsilon}(n)$ $-$ the largest size of the code with error probability $\epsilon$ (assuming equiprobable messages).
A related, commonly used metric is the code rate $R=\log_2{N}/n$.
In this respect, investigating which is the maximum code rate $R_{\epsilon}(n)$ that allows a communication with $n$ channel uses while providing reliability $\epsilon$ is the central question of FBL theory.

Building upon the classical Shannon's asymptotic results, Polyanskiy, Poor, and Verdu  showed that for the additive white Gaussian noise (AWGN) channel with capacity $C (\gamma)={\rm log}_2(1+\gamma)$, where $\gamma$ denotes the signal-to-noise ratio (SNR), the maximal achievable rate with decoding error probability $\epsilon$ can be tightly approximated as~\cite{Polyanskiy}
\begin{equation}
R=C(\gamma)-\sqrt{\frac{V_{\rm AWGN}(\gamma)}{\mathit{n}}}\mathit{Q}^{-1}(\epsilon)+\mathit{O} \left(\frac{\log\mathit{n}}{n}\right),
\label{eq17}
\end{equation}
where $\mathit{Q}^{-1}(\cdot)$ denotes the inverse of the Gaussian Q-function defined as $\mathit{Q}(z)=\int_{z}^{\infty}\frac{1}{\sqrt{2\pi}}\mathit{e}^{-\frac{t^2}{2}}{\rm d}t$, while the so-called channel dispersion $\mathit{V}_{\rm AWGN}(\gamma)$ is defined as
\begin{equation}
\mathit{V}_{\rm AWGN}(\gamma)= \left(1-\frac{1}{1+\gamma^2}\right) {\rm log}^2_2\left(e\right).
\label{D1}
\end{equation}

In the context of this work, we are interested in the error probability of decoding a short-length data packet from a single active user by taking into account the interference contribution from all other active users in a given slot, see (\ref{sinr}). Using the standard ``treat interference as noise'' approximation, one may initially consider the interference as an additional and independent Gaussian noise process.
Strictly speaking, however, the aggregate interference does not obey Gaussian statistics (to achieve \eqref{eq17}, transmitters need to use a non-Gaussian codebook), implying that the OWC receiver experiences non-Gaussian interference. A more precise and appropriate approximation is provided in~\cite{D2}, where non-Gaussian interference and nearest-neighbor decoding are assumed. In this case, the channel dispersion derived in~\cite{D2} (see also~\cite{D1}) should be applied, i.e., 
\begin{equation}
\mathit{V} (\gamma)= \frac{2\gamma}{1+\gamma} {\rm log}^2_2\left(e\right),
\label{D2}
\end{equation}
instead of (\ref{D1}).
On the other hand, for $n>100$ channel uses, the term $O( \frac{\mathrm{log}n}{n})$ in \eqref{eq17} becomes negligible and can be omitted. Taking all this into account, the achievable rate could be tightly approximated as 
\begin{equation}
R \approx C(\gamma)-\sqrt{\frac{V(\gamma)}{\mathit{n}}}\mathit{Q}^{-1}(\epsilon).
\label{eq17b}
\end{equation}
It is to be noted that the approximation \eqref{eq17b} is established by characterizing the asymptotic behavior of analytically tractable achievability and converse bounds (see \cite[Sec. III]{Polyanskiy}). 

For the system setup considered in this paper, relying on equation (\ref{eq17b}) and by setting ${\rm SINR}=\gamma$ (which is dependent on the number of active users $U_a$), we can tightly approximate the decoding error probability as  
\begin{equation}
\epsilon( \gamma, U_a ) = Q \left (\sqrt{\frac{n}{V(\gamma(U_a))}}\left(C (\gamma(U_a))-R\right)\right).
\label{eps}
\end{equation}
For the sake of clarity, in the previous equation we make the dependence of the error probability $\epsilon(\gamma, U_a)$ on the number of active users $U_a$ explicit, although this dependence is implicit through the dependence of the SINR statistics $\gamma(U_a)$ on the number of active users $U_a$.
By conditioning the error probability over the SINR, we get
\begin{equation}
\begin{split}
\epsilon( U_a ) &=\int_\gamma \epsilon( \gamma, U_a ) {f}_{{\rm SINR} } \left( \gamma | U_a \right){\rm d}\gamma ,
\end{split}
\label{eq18}
\end{equation}
where ${f}_{\rm SINR}  \left( \gamma | U_a \right)$ was previously derived in (\ref{pdf_sinr1}).
Finally, the unconditional error probability can be derived as
\begin{align}
    {\epsilon} = \sum_{k=1}^{U} { \epsilon} (  U_a = k  ) \, \mathbb{P} [ U_a = k ],
    \label{Pe}
\end{align}
where
\begin{align}
\label{binom}
\mathbb{P} [ U_a = k ]  = \binom{U}{k} p_a^k ( 1 - p_a)^{U-k}, \; k = 0, \cdots, U,
\end{align}
i.e., the number of active users $U_a$ is a binomial random variable for the assumed Bernoulli arrival process.

\subsection{Throughput for SA via the FBL approach}

The overall system throughput is analysed as the parameter that describes the effective information transmission rate in the system. Based on the system throughput performance, it can be determined if channel uses are effectively exploited, not only due to the effects of the noise but also due to the interference among randomly activated users. 

Considering the FBL regime with the fixed code rate $R$, and the probability that the packet is decoded at each slot as derived previously in (\ref{Pe}), we define the throughput as 
\begin{equation}
\begin{split}
    T & = 0 \cdot \mathbb{P} [ U_a = 0] +  \sum_{k=1}^{U} R \cdot (  1 - \epsilon (  U_a = k  ) ) \cdot \mathbb{P}  [ U_a = k ]  \\
    & = R \cdot ( \mathbb{P} [U_a > 0 ]  - \epsilon) \\
    & = R \cdot ( ( 1 - ( 1 - p_a)^U)  - \epsilon ),
\end{split}
\label{T}
\end{equation}
where $\mathbb{P} [ U_a > 0] = 1-(1-p_a)^U$ represents the probability that at least one user is active during the slot.

\subsection{Outage Probability for SA via the FBL approach}

We also analyze the system reliability in terms of the outage probability of a single transmission.
We consider that the outage  will happen if the overall SINR falls below a predetermined SINR threshold, denoted by $\gamma_{\rm th}$.

For the system under consideration, and under the FBL regime with the selected code rate $R$, a predetermined $\gamma_{\rm th}$ threshold can be defined as a minimal value of the received SINR required to achieve a certain value of the error probability $\epsilon_{\rm th}$. Based on (\ref{eps}), the outage threshold can be determined as
\begin{equation}
\begin{aligned}
\gamma_{\rm th}=\Phi^{-1} (\epsilon_{\rm th}),
\end{aligned}
\label{gTH}
\end{equation}
where $\gamma=\Phi^{-1} (\epsilon)$ is obtained by inverting the expression in equation (\ref{eps}).

Considering the transmission from the reference user, the outage probability can be determined as 
\begin{equation}
\begin{aligned}{ \rm P_{out}} ( U_a ) \! =\!\mathbb{P}\left[{ \rm SINR \!<\! \gamma_{\rm th}} |  U_a \right]\! = \! F_{{\rm SINR}}\left( \gamma_{\rm th}| U_a \right),
\end{aligned}
\label{Pout}
\end{equation}
where $ F_{\rm SINR } \left( \gamma_{\rm th} | U_a \right) $ is in (\ref{cdf_sinr}).
The unconditional outage probability can be calculated as
\begin{align}
    {\rm P_{out}} = \sum_{k = 1}^{U} { \rm P_{out}} (  U_a = k  ) \, \mathbb{P} [ U_a = k ],
    \label{Pout1}
\end{align}
where $\mathbb{P} [ U_a = k ]$ was calculated in \eqref{binom}.
Note that  the reliability can be simply calculated as ${ \rm P_R} = 1 - {\rm P_{out}} $.

\subsection{Special Case: SA without Capture in the FBL Regime}

The classical SA scheme without capture effect assumes that if more than one user is active in a given slot, a collision will happen and the packet is lost. Contrary to SA with capture, the classical SA receiver will only recover the packet if just one of the $U$ IoT devices is active, i.e., $U_a=1$. In that case, $ \epsilon (  U_a = k  ) = 0$ for $k>1$, and thus the error probability is 
\begin{align}
    {\epsilon}  = { \epsilon} (  U_a = 1  ) \, \mathbb{P} [ U_a = 1 ],
    \label{Pe1}
\end{align} 
where ${ \epsilon} (  U_a = 1  )$ and $\mathbb{P} [ U_a = 1 ]$ are determined by expressions (\ref{Pe}) and (\ref{binom}), respectively, where $k=1$.

Based on this, the throughput for SA without capture via the FBL approach can be defined as 
\begin{equation}
\begin{split}
    T & = 0 \cdot \mathbb{P} [ U_a = 0] + R \cdot (  1 - \epsilon (  U_a = 1  ) ) \cdot \mathbb{P}  [ U_a = 1 ] \\
    & +  \sum_{k=2}^{U} R \cdot (  1 - \epsilon (  U_a = k  ) ) \cdot \mathbb{P}  [ U_a = k ]  \\
    & = R \cdot (  \mathbb{P}  [ U_a = 1 ] - \epsilon  )
\end{split}
\label{T1}
\end{equation}
since $ \epsilon (  U_a = k  ) = 0$ for $k>1$.

The outage probability for the OWC system based on SA without capture in the FBL regime can be calculated as 
\begin{align}
    {\rm P_{out}} =  { \rm P_{out}} (  U_a = 1 ) \, \mathbb{P} [ U_a = 1]
    \label{Pout2}
\end{align}
since $ { \rm P_{out}} (  U_a = k ) = 0$ for $k>1$.

Note that there will be no interference contribution under this scenario, since packet can be decoded only if one user is active. According to this, for the error probability derivation in (\ref{eq18}),  the PDF given by (\ref{pdf_hi2}) will be considered instead of the PDF ${f}_{{\rm SINR} } \left( \gamma | U_a \right)$ derived in (\ref{pdf_sinr1}). Similarly, the outage probability will be derived based in the CDF given by (\ref{cdfVLC}) instead of the CDF of the SINR  ${F}_{{\rm SINR} } \left( \gamma | U_a \right)$ derived in (\ref{cdf_sinr}).

\section{Numerical Results and Discussion}

In this section, using the analysis presented in Sec. III and Sec. IV, we focus on the design and performance evaluation of the proposed indoor OWC IoT system.
The results are obtained by adopting the system parameters given in Table I, which represent a typical situation in such environment.
Further, since both the PDF and CDF of the ${\rm SINR}$ in (\ref{pdf_sinr1}) and (\ref{cdf_sinr}), respectively, are given in integral form,  the results are obtained by numerical computations performed through the FFT algorithm implemented in MATLAB.

\begin{table}[b]
\centering
\caption{\bf System parameters}
\begin{tabular}{ccc}
\hline
name  & symbol & value \\
\hline
Transmitted optical power & $ P_t $  & $ 30~\rm{mW}$\\
Photodetector surface area & $ A_r $  & $ 1~{\rm cm}^2 $\\
Responsivity         & $R_r$  & $0.4~{\rm A}/{\rm W}$\\
Optical filter  gain & $T_s$ & $1$ \\
Refractive index of lens at a PD & $\zeta$ & $1.5$ \\
FoV of receiver & $\Psi$ & $90^{\circ}$ \\
Optical-to-electrical conversion efficiency  & $\eta$ & $0.8$ \\
Noise power spectral density & $N_0$ & $10^{-21}~{\rm W}/{\rm Hz}$ \\
System noise bandwidth  & $B$ &  $B=200~{\rm kHz}$\\
\hline
\end{tabular}
  \label{table}
\end{table}

%the OWC IoT device transmitted optical power is $P_t = 30$ mW, the OWC AP receiver PD surface area equals to $A_r=1~{\rm cm}^2$, its responsivity is $R_r=0.4~{\rm A}/{\rm W}$, the FoV of the PD receiver is $90^{\circ}$, while the optical filter gain is $T_s =1$, and the refractive index of the lens at the PD is $\zeta =1.5$. Furthermore, the receiver optical-to-electrical conversion efficiency is $\eta=0.8$, while the noise power spectral density takes a value $N_0=10^{-21}~{\rm W}/{\rm Hz}$, and the system noise bandwidth is chosen to be $B=200~{\rm kHz}$. 
%\CS{This can also go in a table.} \FE{(In fact, in this way it will be more clear. It could also be convenient to stress that the values chosen may well correspond to a hypothetically real setup.)}

Fig.~\ref{FigCDF} presents the CDF of the ${\rm SINR}$ derived in (\ref{cdf_sinr}), comparing the results obtained by the numerical computations with the ones obtained via Monte Carlo (MC) simulations.
Obviously, these results match each other, validating the analysis performed in the previous section.   
In addition, we note that an increase in $U_a$ leads to a situation in which the SINR progressively degrades, which could be expected.

\begin{figure}[t]
\centering
\includegraphics[width=\columnwidth]{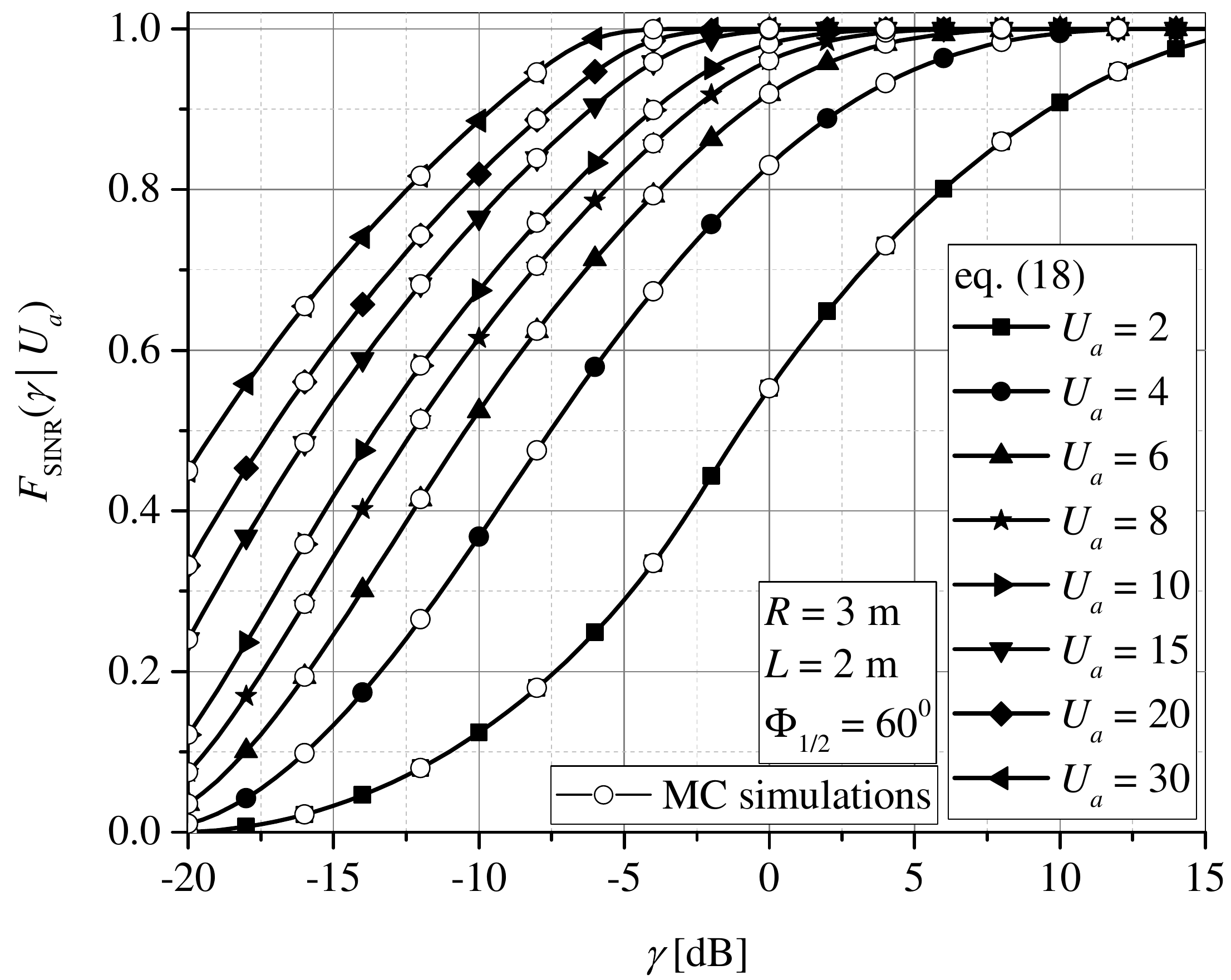}
\caption{SINR CDF of the reference user conditioned on $U_a$, $F_{{\rm SINR}}\left( \gamma_{\rm th} | U_a \right)$. }
\label{FigCDF}
\end{figure}

\begin{figure}[t!]
\centerline{\includegraphics[width=3.5in]{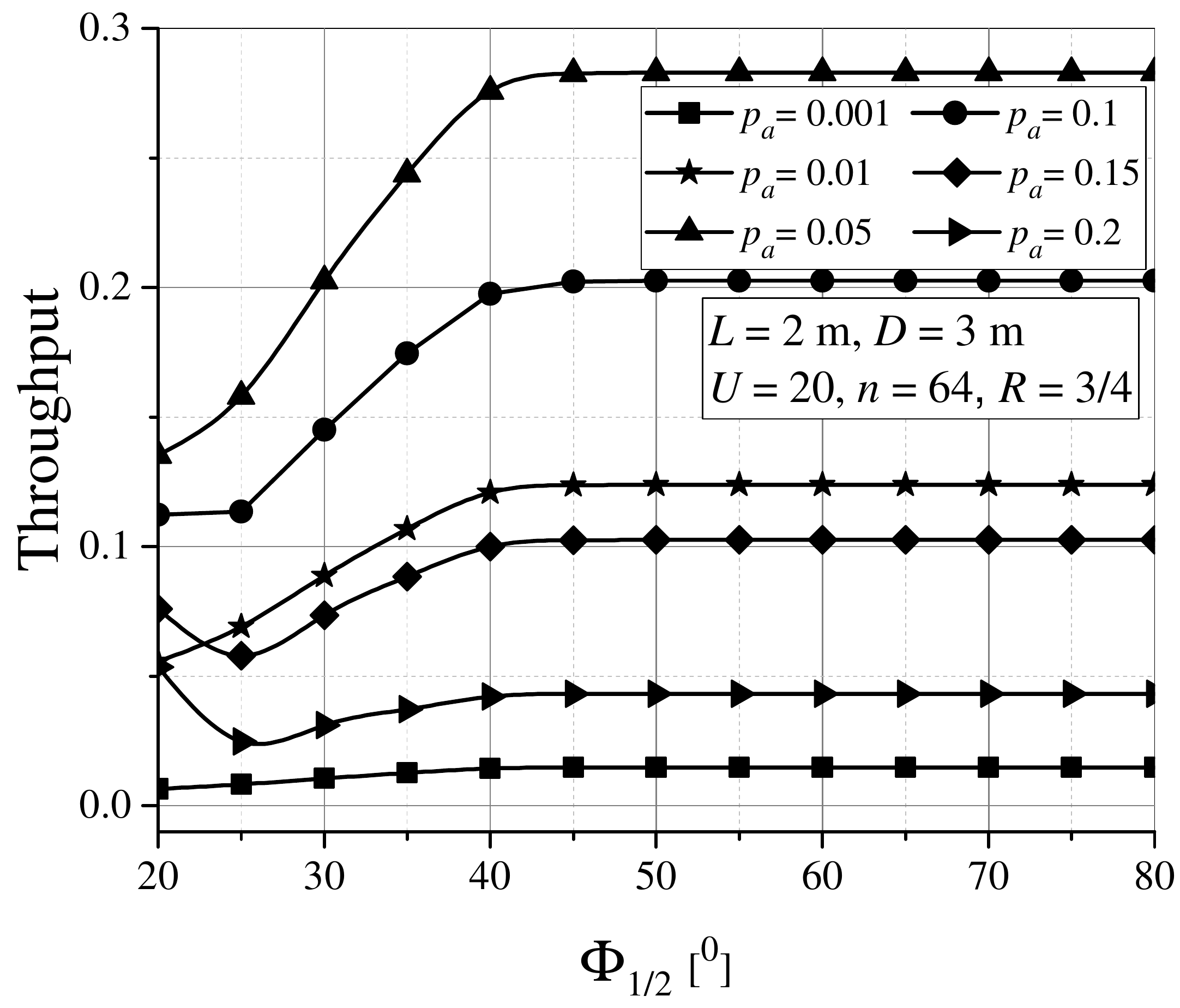}}
\caption{Throughput vs. $\Phi_{1/2}$ for different activation probability values $p_a$. 
%\FE{(For this figure, a number of channel uses $n=64$ is chosen, but we are using (22) in the derivations, which is an approximation valid in theory for $n>100$. I think that this requires an explanation (e.g. justifying that the use of $n$ below that value does not essentially affect the results)}
}
\label{Fig_T_SA}
\end{figure}

\begin{figure}[t!]
\centerline{\includegraphics[width=3.5in]{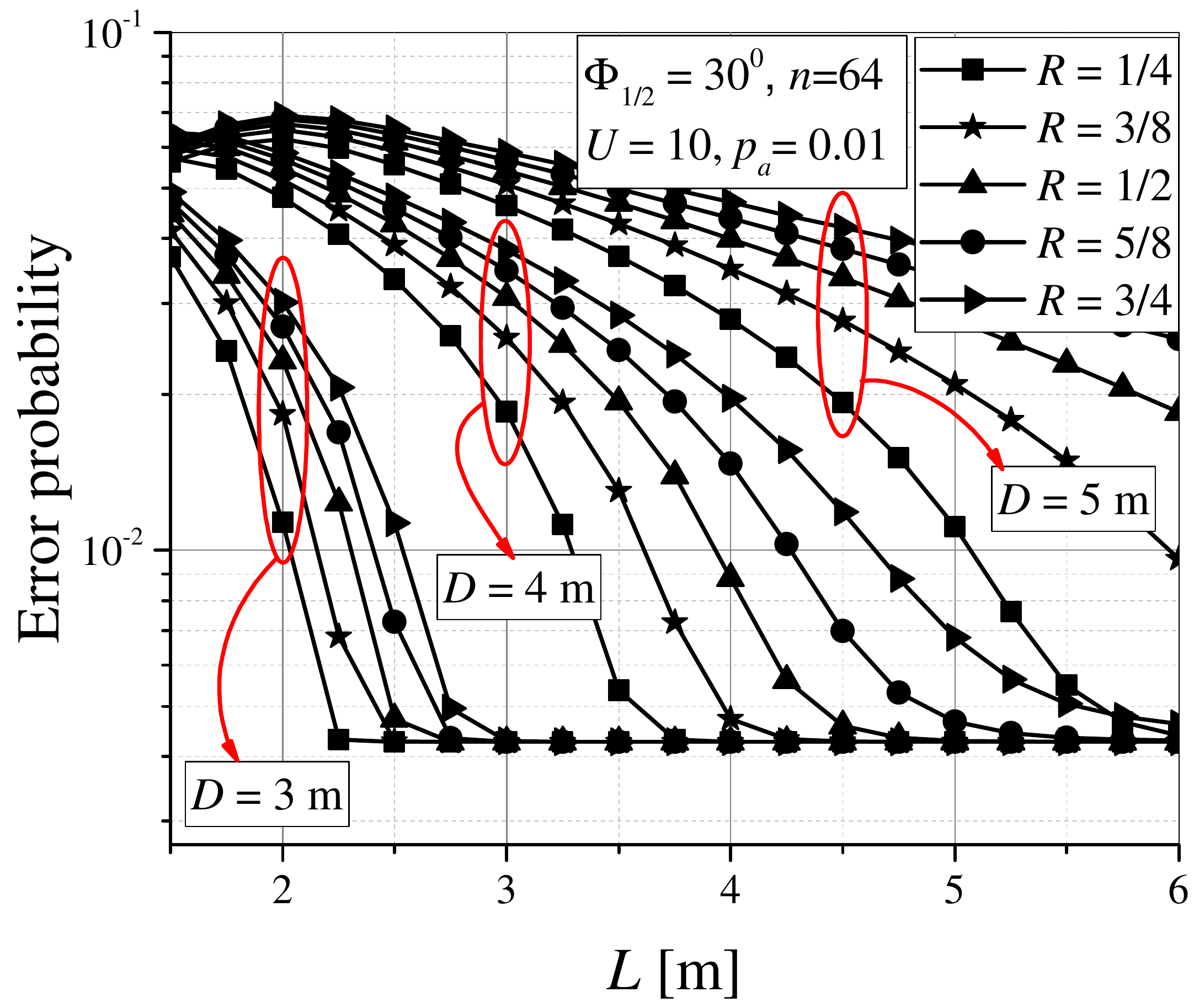}}
\caption{Error probability vs. $L$ for different values of the rate $R$ and  the radius $D$. }
\label{Fig_Pe1}
\end{figure}

Fig.~\ref{Fig_T_SA} depicts the dependence of the throughput on the semi-angle at half illuminance $\Phi_{1/2}$, evaluated using the expression derived in (\ref{T}), and considering different values of the activation probability $p_a$.
An increase in $p_a$ up to a certain value results in a better performing system from the throughput point of view.
However, after this value of $p_a$ is surpassed, the throughput starts to decrease.
In fact, it can be seen that there is an optimal value of $p_a$ that maximizes the throughput performance of the system.
%Furthermore, Fig.~\ref{Fig_T_SA} shows how the throughput behaves for different values of the semi-angle $\Phi_{1/2}$.
On the other hand, depending on $p_a$ (i.e., the number of active OWC-based IoT devices), a higher value of $\Phi_{1/2}$ may improve or degrade the system performance.
The semi-angle at half illuminance of the LED determines the Lambertian order of the LED source, or, in other words, determines the received optical signal intensity from the active IoT users at the OWC AP.
When the value of $\Phi_{1/2}$ is higher, the optical beam at the LED source output will be wider, thus the received power from the users that are not close from the AP will be higher.
This holds for all IoT devices, thus the semi-angle will have a strong impact on both the reference user and the interference contributions (i.e., on the overall received SINR).
To conclude, since $\Phi_{1/2}$ represents an important parameter which has a substantial impact on the received optical signal intensity, the number of active users (together with the overall interference) will determine if an increase in $\Phi_{1/2}$ leads to a higher or a lower system throughput.

Fig.~\ref{Fig_Pe1} depicts the error probability of the considered OWC system in the FBL regime, as derived in (\ref{Pe}), as a function of the OWC AP height $L$, for various levels of error protection parameterized by the code rate $R$.
Additionally, three different values of radius $D$ are considered. 
As expected, increasing $R$ will result in an error protection performance deterioration. 
When the radius $D$ is larger, the OWC IoT users will be distributed over a wider area, which will increase the chance that many of them are far distant from the OWC AP.
Increasing the distance of the OWC users from the AP leads to a reduced received power, and consequently the overall SINR will be lower.
This will affect the capture probability, and thus the error probability of the system will increase when $D$ increases.
%Since the distances of the OWC users from the AP are also dependent on the height $L$, \CS{the error probability is reduced to certain point with increase of  $L$. - not sure what is really meant here, try to be more precise}
Further,  Fig.~\ref{Fig_Pe1} shows that there is an error probability floor, % from a given height value, %as a function of the rest of parameters. 
which is reached for lower values of $L$ %for the error probability floor gets lower 
as the radius $D$ gets smaller.
When the users are located in a wider area, the error floor appears at very high $L$.
Note that the error probability floor is independent on the code rate $R$, as well as on the radius $D$.
In other words, increasing the distance between the planes where the IoT users and the OWC AP are placed will not result in lower-error probability after the error floor is reached, for any value of $D$ and $R$. 
Consequently, the geometric setup of the OWC based IoT framework has an impact on the system performance behavior in the FBL regime. %, but there are parameters that affect the results more or less dramatically (for example, the mentioned lack of sensitivity of the error probability floor level with $L$).

\begin{figure}[t!]
\centerline{\includegraphics[width=3.5in]{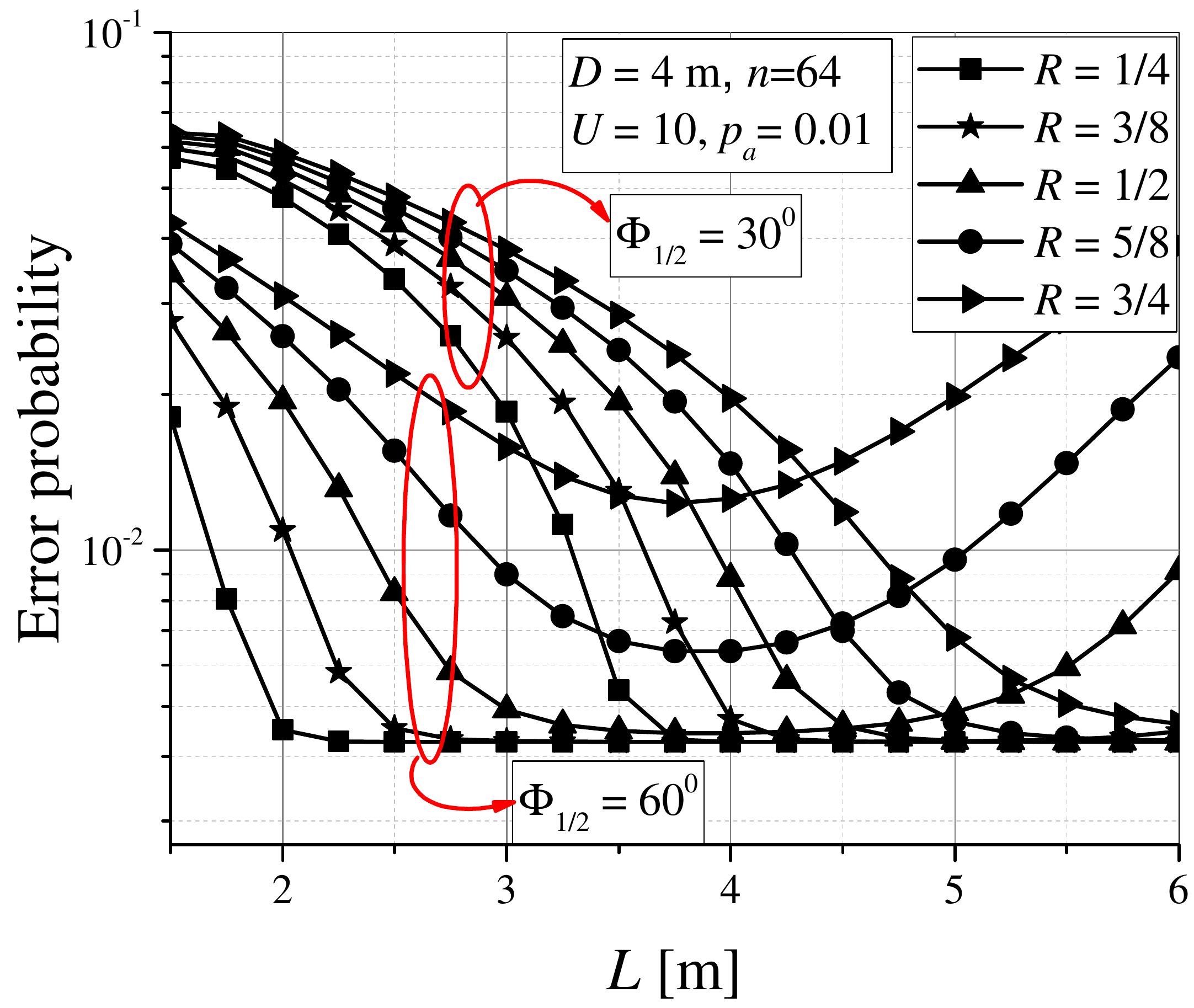}}
\caption{Error probability vs. $L$ for different values of the rate $R$ and the semi-angle $\Phi_{1/2}$.}
\label{Fig_Pe2}
\end{figure}

Fig.~\ref{Fig_Pe2} presents the error probability $\epsilon$ of the considered OWC system under the FBL regime, derived in (\ref{Pe}), as a function the OWC AP height $L$, for different code rates $R$.
In contrast to Fig.~\ref{Fig_Pe1}, the system behavior here is observed for two different values of the semi-angle at half illuminance $\Phi_{1/2}$. Similarly to Fig.~\ref{Fig_T_SA}, the system behaves differently when the value of $\Phi_{1/2}$ increases.
For example, when $L<3$ m, the error probability is lower for the larger semi-angle $\Phi_{1/2}$ value, whereas for $L>5$ m, the system performance degrades sharply for higher values of $\Phi_{1/2}$.
Additionally, a larger distance from the OWC IoT users to the AP (i.e., greater $L$) leads to a reduced received power, and to a lower overall SINR. This will affect the capture probability in such a way that larger $L$ results in a behavior where the error probability of the system considered will initially decrease (eventually reaching the error probability floor in some cases), and, after some point that is dependent on $\Phi_{1/2}$, it will increase.
Moreover, there is a minimum value of the error rate for different OWC system setups, determined by the semi-angle $\Phi_{1/2}$ and the overall distance from the OWC IoT users to the AP.
From Figs.\ref{Fig_Pe1} and~\ref{Fig_Pe2}, it can be concluded that the effects of the rate value on the error probability performance is mainly dependent on the parameters $L$ and $\Phi_{1/2}$.
% The system performance behavior under the FBL regime is highly dependent on the geometric setup of the OWC based IoT system. 

\begin{figure}[t!]
\centerline{\includegraphics[width=3.5in]{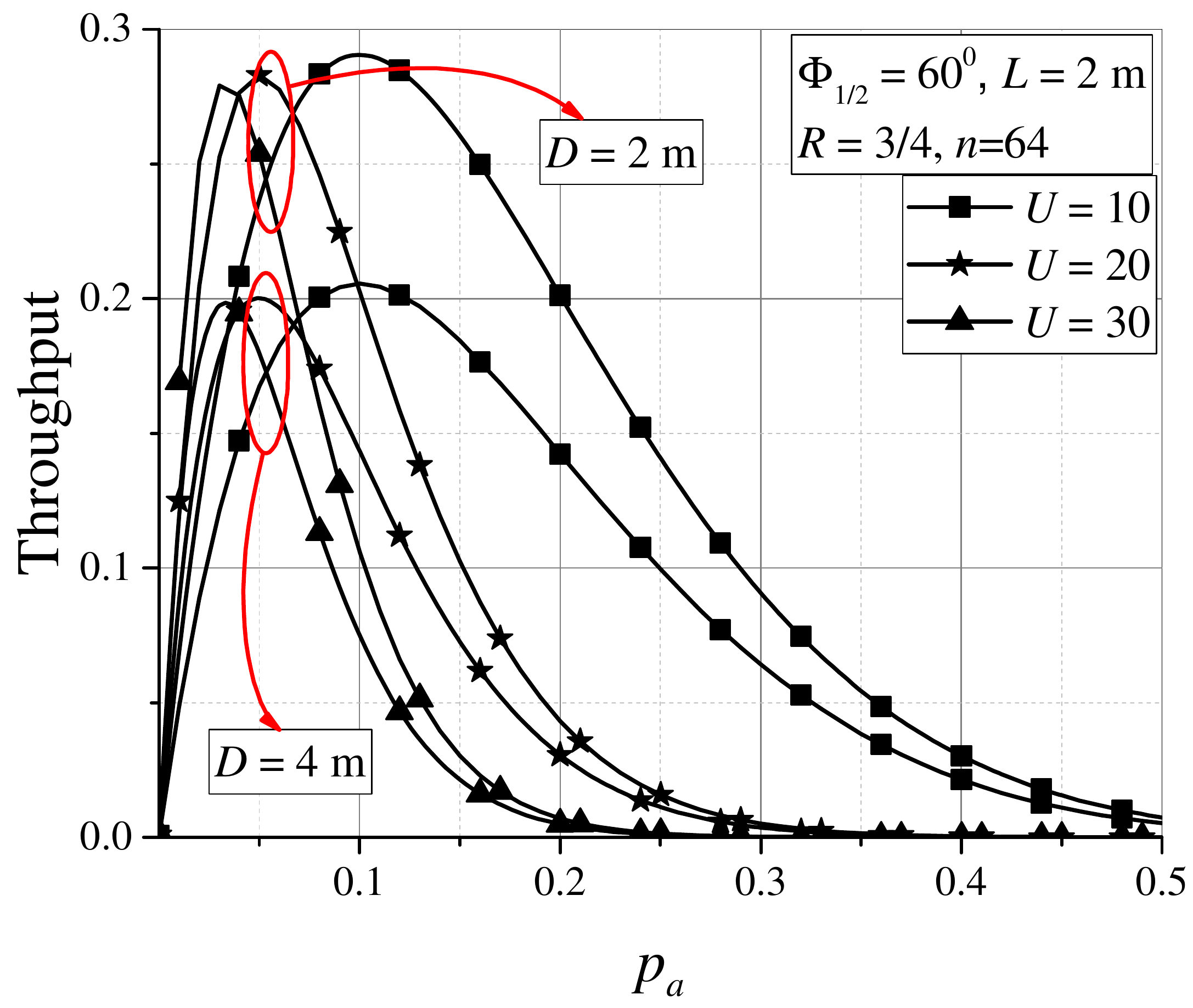}}
\caption{Throughput vs. $p_a$ for different number of total IoT devices $U$.}
\label{Fig_T_D}
\end{figure}

\begin{figure}[t!]
\centerline{\includegraphics[width=3.5in]{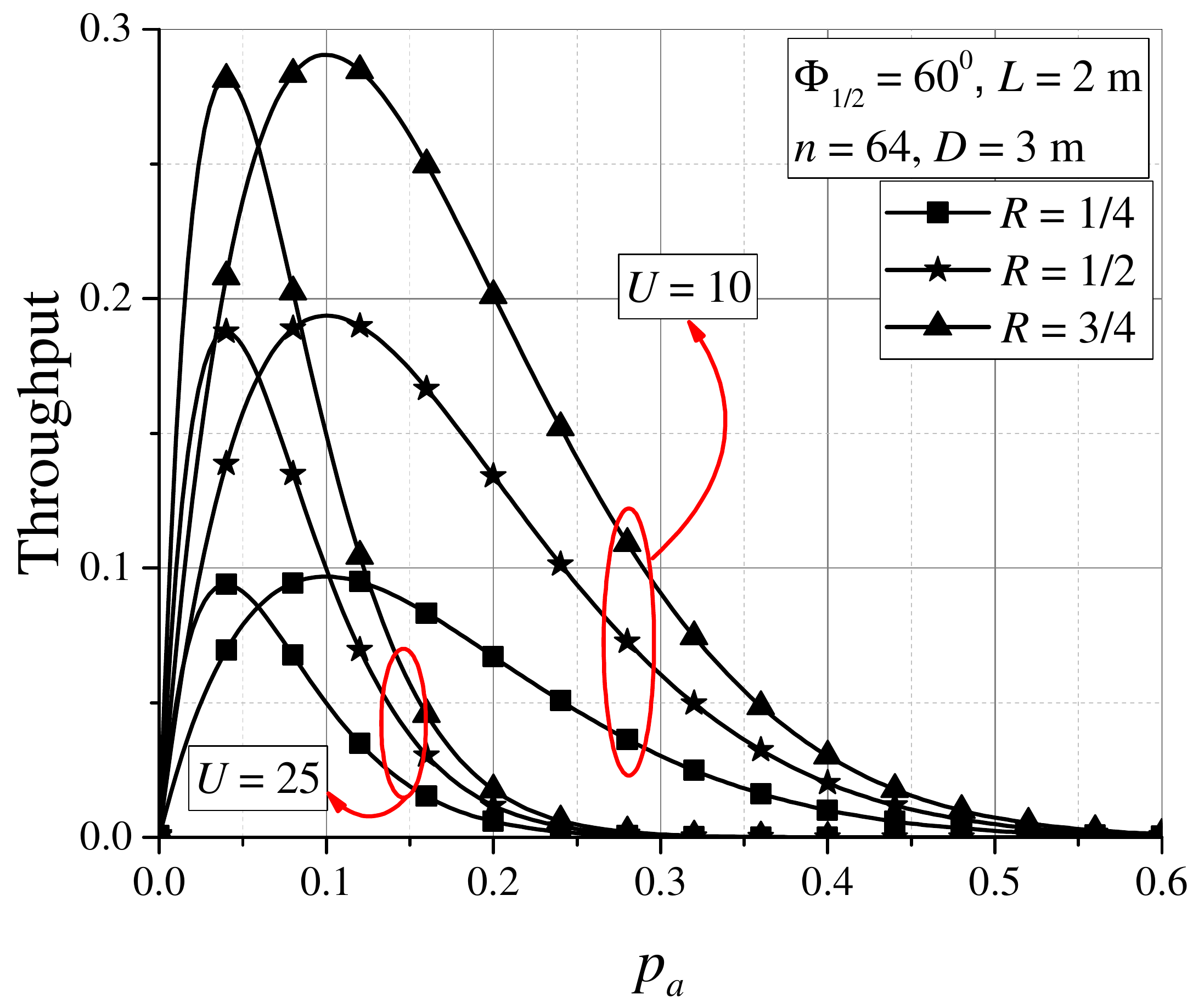}}
\caption{Throughput vs. $p_a$ for different values of the maximum achievable rate $R$.}
\label{Fig_T_R}
\end{figure}

Fig.~\ref{Fig_T_D} shows the dependence of the throughput derived in (\ref{T}) on the activation probability $p_a$ for different number of OWC-based IoT devices $U$ and different values of radius $D$.
As previously stated, a higher value of the radius $D$ leads to a worse system performance, i.e., to a lower throughput level.
A higher activation probability $p_a$ and a higher number of total users $U$ implies a higher number of active users in a slot. The number of active IoT devices determines the overall SINR, which has a direct impact on the potential of the capture effect. 
Consequently, as for the trend of each curve, it can be seen how, when the total number of users $U$ in the system gets higher, the optimal value of $p_a$ that maximizes $T$ gets lower.
Moreover, after achieving its maximum value, the system throughput starts decreasing with a further increase of $p_a$.
Given the radius $D$, the maximum value of the throughput is rather insensitive the number of users. %, but it differs when changing the radius $D$.
On the other hand, the optimal value of $p_a$ that maximizes the throughput depends on the total number of users $U$, but, given $U$ it remains the same irrespective to the change in $D$.
%but remains the same when changing the radius $D$.
In conclusion, we can say that the geometric setup of the OWC based IoT system and the total number of users have strong impact on the optimal system performance, and should be taken into consideration during the design of the access protocol.

In Fig.~\ref{Fig_T_R}, we show the throughput derived in (\ref{T}) as a function of the activation probability $p_a$ for different values of the maximum achievable rate $R$.
As previously stated, a higher rate corresponds to a degraded system performance.
Moreover, Fig.~\ref{Fig_T_R} also reveals that there is an optimal value of $p_a$ that leads to a maximum value of the system throughput. From the presented results, it can be concluded that the optimal value of $p_a$ differs for different number of overall users, but it is the same regardless of the employed rate $R$. % in this FBL based OWC scenario.

\begin{figure}[t!]
\centerline{\includegraphics[width=3.5in]{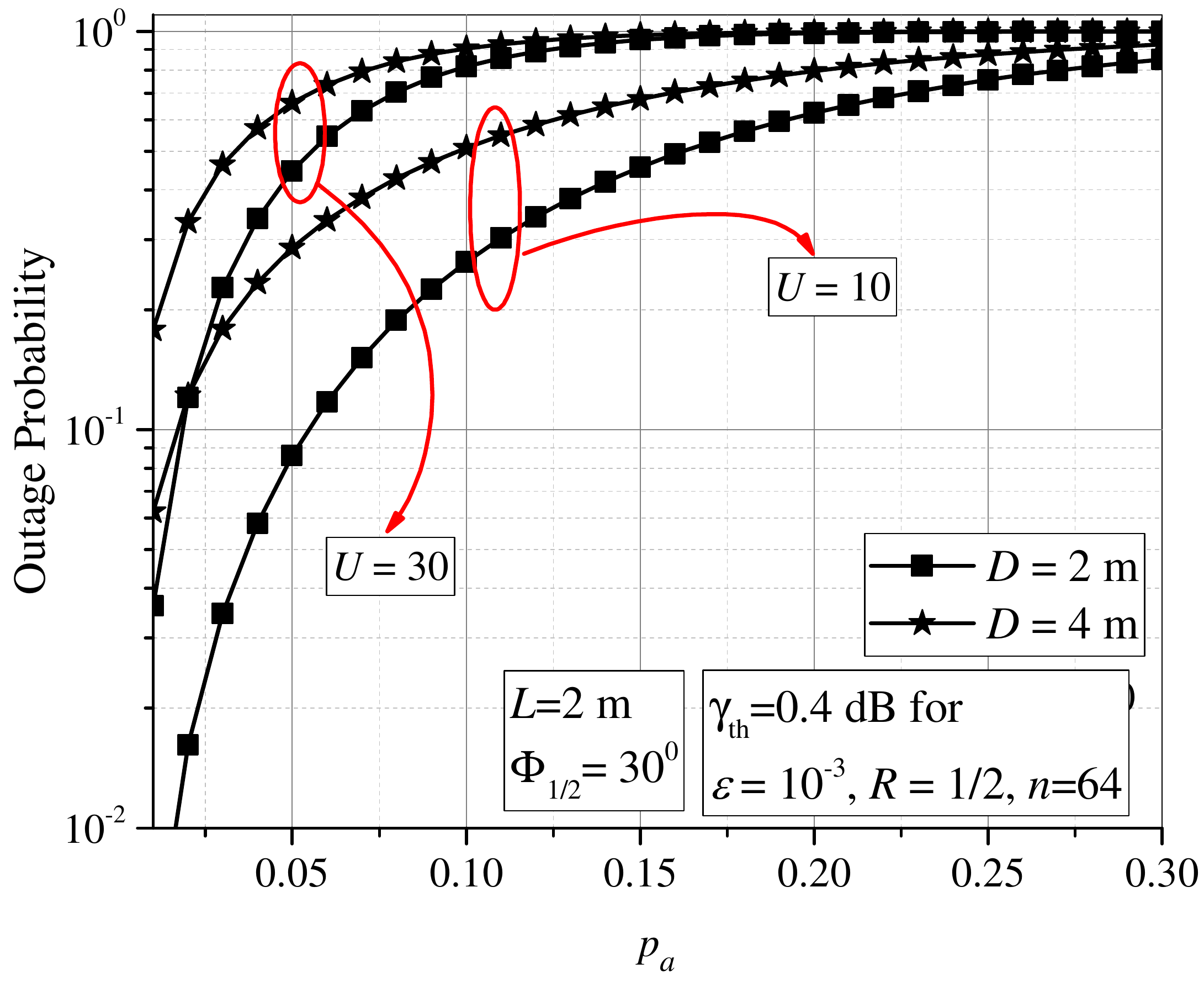}}
\caption{Outage probability vs. $p_a$ for different values of the radius $D$ and different number of users $U$.}
\label{Fig_Pout}
\end{figure}

Based on the expression derived in (\ref{Pout1}), Fig.~\ref{Fig_Pout} depicts the outage probability dependence on the activation probability $p_a$, considering two values for the radius $D$ ($D=2$~m and $D=4$~m).
The outage threshold $\gamma_{\rm th}$ was defined in (\ref{gTH}), and is obtained for an error probability of $\epsilon_{\rm th}=10^{-3}$, a rate $R=1/2$ and a number of channel uses $n=64$. 
As previously stated, a smaller value of the radius $D$ results in a better system performance.
As shown in the figure, a higher number of users lead to a system performance deterioration, since the interference contribution will be stronger.
Furthermore, it can be noticed that the radius $D$ and the total number of users $U$ have an effect on the outage performance only for a smaller number of active users, while when $p_a>0.25$ the radius $D$ and the number $U$ seem to have no impact on the outage probability.
More precisely, the system outage will happen for $p_a>0.3$, i.e., $P_{\rm out} \approx 1 $, for any $D$ and $U$, since the overall SINR will not be higher than $\gamma_{\rm th}$ and the error probability target value $\epsilon_{\rm th}=10^{-3}$ will not be achieved. 

Figs.~\ref{Fig_comparison1} and \ref{Fig_comparison2} present the overall throughput comparison of the proposed SA OWC IoT system with capture (eq. (\ref{T})), and the SA system without capture effect (eq. (\ref{T1})), in the FBL regime. 

Fig.~\ref{Fig_comparison1} shows the throughput dependence on $p_a$ for various levels of error protection parameterized by the code rate $R$.
A significant improvement of the system throughput is noticed in presence of the capture effect, especially for large code rates $R$.
The results prove that the designed RA protocol represents an improvement from the point of view of the throughput, since user data can be recovered even in presence of concurrent transmissions thanks to the successful exploitation of the capture effect.

\begin{figure}[t!]
\centerline{\includegraphics[width=3.5in]{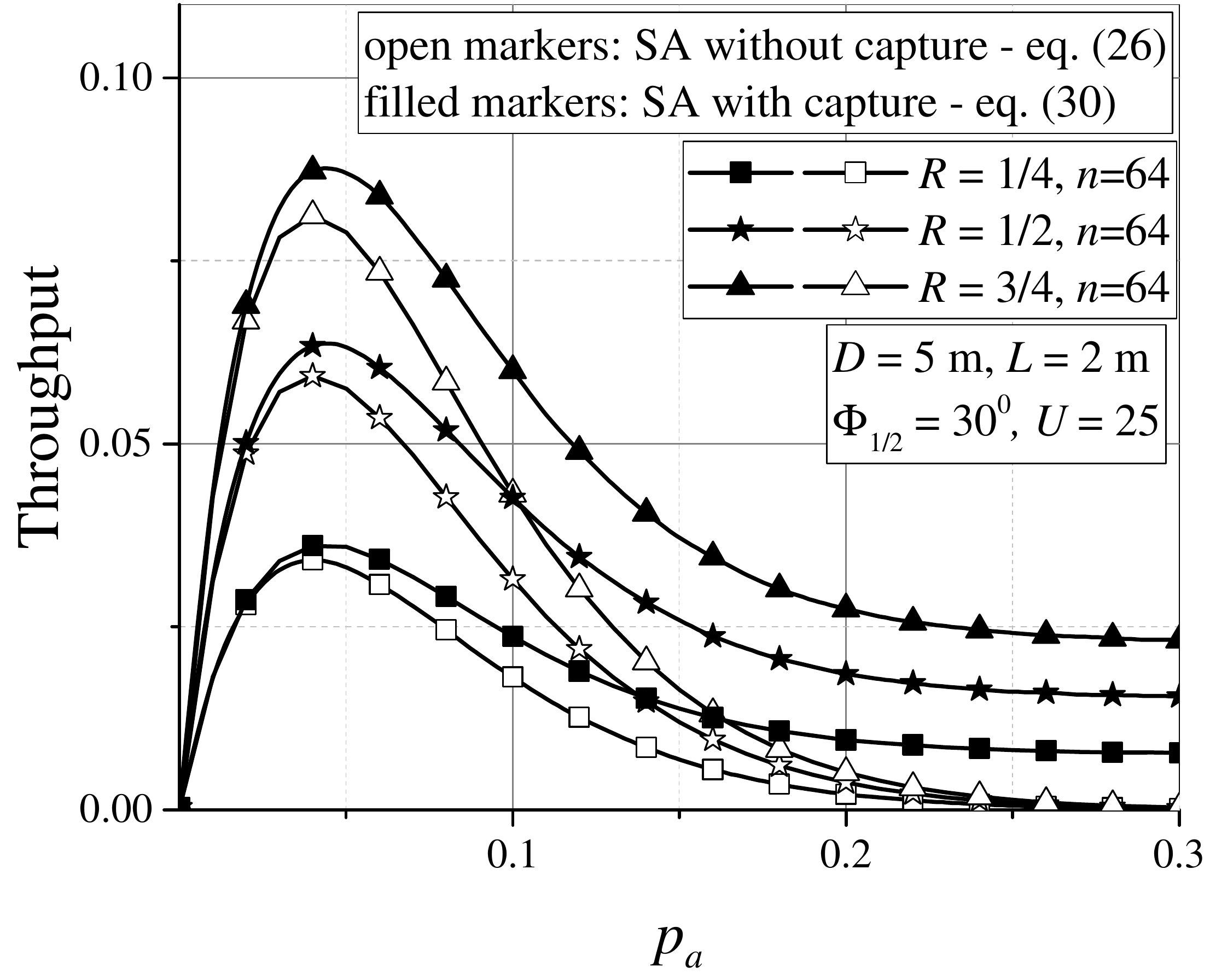}}
\caption{Comparisons of the throughput of the system with and without capture effect for different values of the rate $R$.}
\label{Fig_comparison1}
\end{figure}

Fig.~\ref{Fig_comparison2} depicts the throughput dependence on $p_a$ for two values of the semi-angle $\Phi_{1/2}$. The throughput improvement achieved by employing SA with capture is only evident for the lower semi-angle of $\Phi_{1/2}=30^0$. As a contrast, the throughput remains the same when the semi-angle is higher, i.e., $\Phi_{1/2}=60^0$. It can be concluded that the semi-angle of the OWC system plays a crucial role with respect to the interference contribution and its effects. More specifically, the capture effect leads to significant throughput improvement under the specific condition of limited interference (i.e. lower semi-angle), as may be expected. 

In summary, we can conclude that the geometric setup (including the semi-angle at half illuminance $\Phi_{1/2}$, the radius $D$ and the height $L$), the number of active users, and %the FBL regime parameterized by 
the code rate $R$ play an important role in the possible maximization of the system performance.
These parameters should be taken into account %and tackled according to the principles illustrated during the design 
in order to appropriately design an access scheme in an OWC-based IoT system.

\begin{figure}[t!]
\centerline{\includegraphics[width=3.5in]{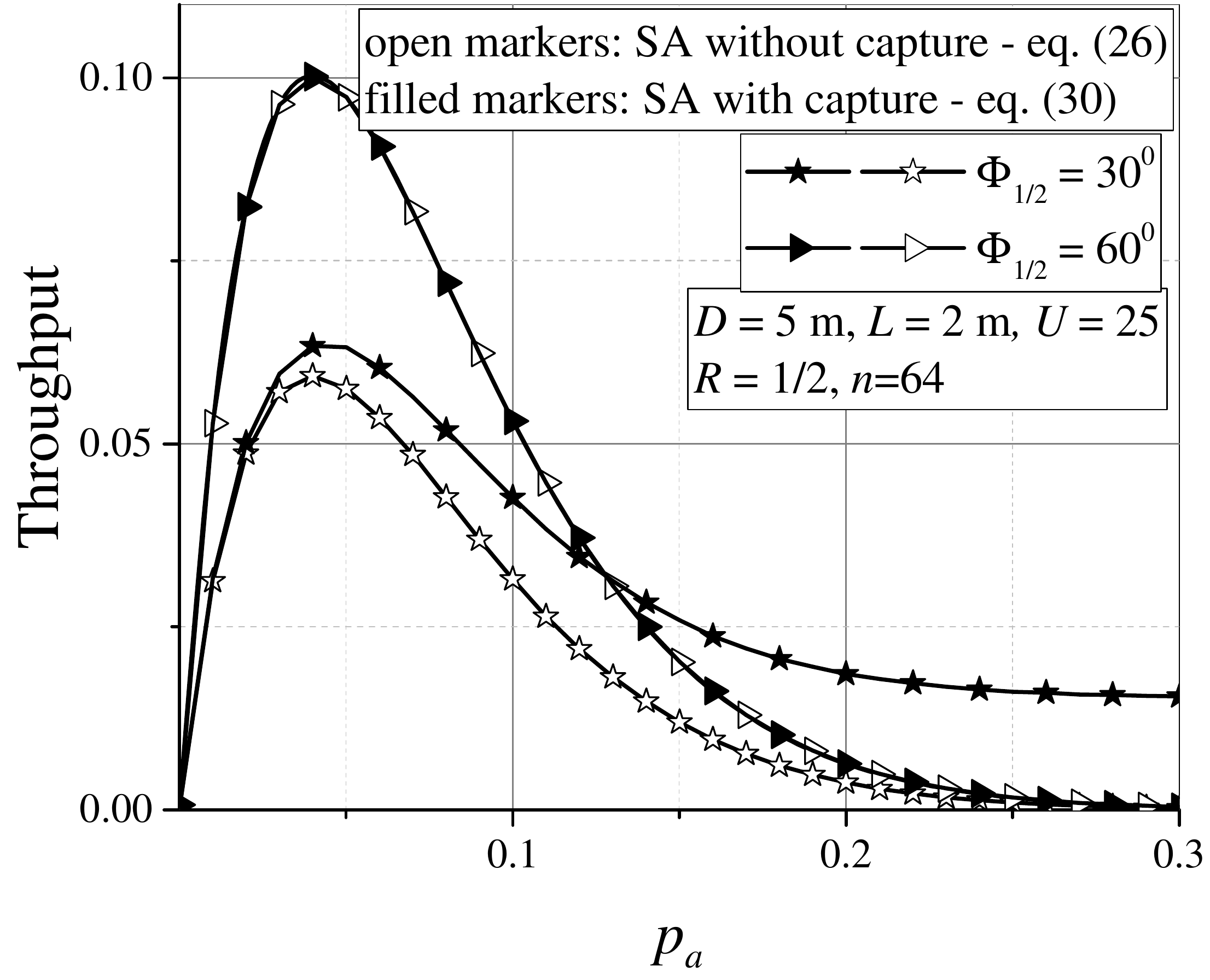}}
\caption{Comparisons of the throughput of the system with and without capture effect for different values of the semi-angle $\Phi_{1/2}$.}
\label{Fig_comparison2}
\end{figure}

\section{Conclusion}

In this paper, we have analyzed the uplink of an indoor OWC-based IoT system under the FBL regime and with a single access point.
The system employs a SA-based access scheme, which exploits the capture effect, i.e., the possibility to decode user transmissions in the presence of interference from other users, which inherently exists in such setups.
%meaning that the collisions introduced by the interference contribution have been taken into account while determining the system performance.
Based on the received SINR statistics, an analytical expression for the error probability in the FBL regime has been derived, as well as an expression for the overall throughput and the outage probability of the system under investigation.
The derived expressions are used to obtain numerical results, which are further analyzed to assess the system performance behavior depending on the OWC framework and the channel model parameters.

The results presented have shown how the geometric setup of the OWC-based IoT system  and the activation probability affect the error probability, the throughput and the outage probability performance. It has been proven that the OWC-based system geometry has a significant impact on the overall system performance, and should be taken into account during the design of any indoor OWC-based IoT system in order to assess and optimize the RA protocol performance.

\section*{Acknowledgment}
This work has received funding from the European Union Horizon 2020 research and innovation programme under the grant agreement No 856967 and by the Secretariat for Higher Education and Scientific Research of the Autonomous Province of Vojvodina through the project “Visible light technologies for indoor sensing, localization and communication in smart buildings” (142-451-2686/2021). This research was supported by the Science Fund of the Republic of Serbia, Program DIASPORA, \#GRANT 6393139, AVIoTION and by the European Union COST Action CA19111.

\end{document}